\documentstyle[numreferences]{klu}

\def\phiB {\mbox{${\phi \kern-.582em {\phi} \kern-.576em {\phi}}$}}
\def\psiB {\mbox{${\psi \kern-.662em {\psi}  \kern-.666em {\psi}}$}}

\begin{opening}
\title{Hyperbolic Models of Homogeneous Two-Fluid Mixtures}
\author{S.L. \surname{Gavrilyuk}}
\author{H. \surname{Gouin}}
\institute{ Laboratoire de Mod\'elisation en
M\'ecanique et Thermodynamique, Universit\'e d'Aix-Marseille, Facult\'e des
Sciences et Techniques, Case 322, Avenue Normandie-Niemen, 13397 Marseille Cedex
20, FRANCE} \author{Yu.V. \surname{Perepechko}}
\institute{United Institute of Geology, Geophysics and Mineralogy,
Siberian Branch of the Russian Academy of Sciences,
630090 Novosibirsk, RUSSIA}
\date{}
\end{opening}

\begin{document}

{\abstract{
One derives the governing equations  and the Rankine - Hugoniot conditions
for a mixture  of two miscible fluids  using  an extended form of  Hamilton's
principle  of least action.
The Lagrangian is constructed as the difference between the kinetic
energy and a potential depending on the relative velocity
of components. To obtain the governing equations and the jump conditions
one uses two reference frames related with the Lagrangian
coordinates of each component.
Under some hypotheses on flow properties one proves
the hyperbolicity  of the governing system for small relative velocity of
phases.}}

\bigskip

\noindent
{\bf Sommario.} \ Le equazioni di governo e le condizioni di
Rankine-Hugoniot sono derivate per una miscela di due fluidi
miscibili usando una formulazione estesa del principio di minima
azione di Hamilton. La Lagrangiana {\`e} costruita come differenza tra
energia cinetica e
potenziale dipendente dalla velocit{\`a} relativa dei componenti. Per
ottenere le equazioni di governo e le condizioni di salto si usano
due sistemi di riferimento collegati alle coordinate Lagrangiane di
ciascun componente. Nelle stesse ipotesi sulle propret{\`a}
di flusso si prova l'iperbolicit{\`a} del sistema per piccole velocit{\`a}
relative delle fasi.

\bigskip

{\keywords{Hamilton's principle, Hyperbolicity, Multiphase Flows.}}

\section{Introduction}

 The theory of mixtures considers two different kinds of continua:
 {\it homogeneous}
 mixtures (each
 component occupies the whole volume of the physical space) and
 {\it heterogeneous} ones (each component
 occupies only a part of the mixture volume). In
 the second case, the geometrical parameters appear as the
 unknown variables: the volume
 concentration, sizes of dispersed particles etc. The problem, which is common for
 both types of media, is to describe two-velocity effects, that are responsible for
 the development
 of instability in mixtures, diffusion of components
 etc.

At least three approaches  to the construction of  two-fluid models are known. The most common one
for studying of  heterogeneous
two-phase flows  is the  {\it averaging method}  (Ishii [1],
 Nigmatulin [2] and others). Averaged equations of
 motion are obtained by applying an
 appropriate averaging operator  to the balances laws
 of mass, energy etc., which are valid inside each
 phase.
 The main problem
 associated with this approach is the closure of the system obtained: the
 system
 contains more unknowns than equations. Different experimental and theoretical
 hypothesis are
 used for the closure. Not all of them  give {\it well-posed}  governing
equations.
 For example, it was noted by many
 authors (Drew [3], Kraiko
\&  Sternin [4], Stuhmiller [5]  and others) that
 non-dissipative averaging governing equations of heterogeneous
two-velocity media are
  {\it not  hyperbolic}  even for small relative velocity of the mixture components,
when the
 equal  pressures  hypothesis in phases is used.  This
 implies that
 the Cauchy problem for the corresponding equations of motion is {\it ill-posed}.
  The nonhyperbolicity was overcome later by many
 authors: Liapidevskii [6],
 Ransom \& Hicks [7]
 (two-phase separated flow), Biesheuvel \&
 van Wijngaarden [8] (bubbly
 liquids), Fabre {\it et al} [9] (slug flow) etc.
 The well-posedness of governing equations was achieved  by using
  additional
 closure relations for averaged quantities, which are specific to the flow
 type. The hyperbolicity of one-dimensional models only
 was proved.

 A different  approach known as {\it Landau   method  of conservation laws}
was initially used for constructing
  models of quantum liquids such as superfluid helium (Khalatnikov [10], Landau \&
 Lifshits [11],  Putterman [12]).
 The method consists  in the following: the requirement of the fullfillment of
balance laws of mass,
 energy etc., complemented by the Galilean relativity principle  and the Gibbs thermodynamic
identity  fully determines the
governing equations
 of motion. Recently this approach was applied to classical fluids
(two-velocity
 hydrodynamics) by Dorovsky \&
 Perepechko [13],  Roberts \&
  Loper [14],   Shugrin [15].
 The method does not take into account the
 geometrical characteristics of the mixture  components: the volume
 concentrations, sizes of
 particles etc. In the non-dissipative case it gives also hyperbolic models
(see, for instance,
Khalatnikov [10], where sound velocities for superfluid helium are calculated).

 Finally, a third  approach called  {\it variational  method}  is the most
universal.  Bedford \&  Drum\-hel\-ler [16],
 Berdichevsky [17],  Geurst [18,19] have applied  it for investigation of bubbly
 liquids flows. In particular,  Geurst
 has proven in
 one-dimensional case the hyperbolicity of the governing equations for
small relative velocity  of phases.

These methods present three different approaches for description of
complex media. At present, their common features  and distinctions  are
not quite well understood.

We consider the variational approach to describe two-velocity effects in homogeneous
 mixtures. A physical example of such a flow is a mixture of two  miscible fluids,
or  a mixture of two gases with quite
 different molecular weights.

In Section 2 we introduce an extended form of Hamilton's principle of least action.
 The Lagrangian
 of the system is chosen in a general form: it is the difference of
 the kinetic energy of the system,
 which depends obviously on the choice of a reference
 frame, and a thermodynamic potential, which is a
 Galilean invariant, conjugated to
 the internal energy  with respect to the relative velocity
 of phases.
 If it does not depend on the relative velocity of components, we have a classical
form of
 Hamilton's action for two-velocity systems (see, for instance, the article
 by Gouin [20], where the thermo-capillary mixtures were studied).

In Section 3 we get from the variational principle the governing equations and
 the Rankine-Hugoniot  conditions for shocks. To obtain the desired relations,
 we used
two reference frames related
 with Lagrangian coordinates of each component.

Conservation laws of the total energy and the total momentum are derived in Section
4. We show that,
 under some restrictions on the flow properties, the governing
 equations admit  additional conservation
 laws in terms of the densities \
$\rho_1$,\, $\rho_2$\ and the velocities \ ${\bf u}_1$,\, ${\bf u}_2$ \ of
components. Without these restrictions
 the system seems not to be conservative. We extend the set of the unknown
variables, considering the
deformation gradients as the required
 quantities, and rewrite our system in a conservative form that gives
 additional set of possible jump conditions.

In Section 5,  classification of strong discontinuities is done and some
 difficulties of the
"right" choice of jump conditions are discussed.

  We investigate in Section 6 the
hyperbolicity of the governing system for small relative
velocity of phases in multi-dimensional case. Under some hypotheses
 on flow properties,
we reduce our system to
Friedrichs' symmetric form and prove that
  convexity of the internal energy
 guarantees
hyperbolicity of the governing equations.

As a convention, in the following we shall use asterisk "*" to denote
{\it conjugate}
(or {\it transpose}) mappings
 or covectors (line vectors). For any vector
\ ${\bf a}$, ${\bf b}$ \ we shall use the notation \ ${\bf a}^* {\bf b}$\  for
their
 {\it scalar  product}
(the line vector is multiplied by the
column vector) and
\ ${\bf a} {\bf b}^*$\  for their
{\it tensor \ product}
 (the column vector is multiplied by the line vector).
The product of a mapping \ $A$ \ by a vector \ ${\bf a}$\ will be denoted by \
$A\, < {\bf a} >$. The notation \ ${\bf b}^* \, A$ \ means
 the covector
\ ${\bf c}^* $ \ defined by the rule \ ${\bf c}^* = ( A^*\, < {\bf b} > )^*$.
The divergence of a linear transformation \ $A$ \ is the covector \ $div A$
such that, for
any constant vector \ ${\bf a}$
$$
div\, A\, {\bf a} \  = \ div \ (A\, < {\bf a} > ).
$$
 The letter \ $I$ \ will mean the identity
transformation, and \ $\nabla$ \ will mean the gradient line operator.
The greek indices \ $\alpha, \, \beta \ = 1, \, 2$ \ will denote the parameters of
components such as the densities \ $\rho_{\alpha}$, the velocities
\ ${\bf u_{\alpha}}$ \ etc.

\section{Variational Principle}

  Let us suppose
 that a  mixture  of two miscible  fluids is well described by the velocities
\ ${\bf u}_1$, \, ${\bf u}_2$ \ of two components, the average
 densities \ $\rho_1$, \, $\rho_2$ \ and the total
energy \ $E$. The total energy is divided into the
 kinetic energy \ $T$ \ and the
internal energy \ $U$.
In the following, we will consider only mechanical processes by
suppressing thermal evolution. Hence, $U$ is purely mechanical part of the total
internal energy.
The kinetic energy, depending on the choice of a reference
frame, is represented by the classic formula:
$$
T  = \ {1\over 2 } \ \rho_1 \vert {\bf u}_1 \vert^2 \ +
{1\over 2 } \  \rho_2 \ \vert {\bf u}_2 \vert^2.
$$
The internal energy \ $U$ \ is a Galilean invariant, it does not depend on the
reference frame. Hence, \ $U$ \ depends
 on  \ $\rho_1$,\, $\rho_2$ \ and the relative
velocity \ ${\bf w}={\bf u}_2 - {\bf u}_1$.

Neglecting   the dissipative effects, we propose the following extended form
of Hamilton's principle of
least action ( Gavrilyuk {\it et al} [21]):
$$
\delta {\cal I} = 0 , \ \ {\cal I} = \int^{t_2}_{t_1} \!\! \int_{\cal D} \ L
\ d {\bf x} d t, \ \ \ \
L \ = \ \rho_1 \ {\vert {\bf u}_1 \vert^2\over 2 } +
 \rho_2 \ {\vert {\bf u}_2 \vert^2\over 2}
\ - W (\rho_1 , \rho_2 , {\bf w})
\eqno(2.1)
$$
with additional kinematic contraints
$$
{\partial \rho_1\over \partial t}+ div \ (\rho_1 {\bf u}_1) = 0 \ , \
{\partial \rho_2\over \partial t}+ div \ (\rho_2 {\bf u}_2) = 0.
\eqno(2.2)
$$
Here \ $\lbrack t_1 , t_2 \rbrack$ \ is a time interval, \ ${\cal D}$ \ is a
fixed bounded domain of the three-dimensional space with the
 boundary \ $\partial {\cal D}$. We will suppose that  the
slipping condition on \ $\partial {\cal D}$ is fulfilled  for each component.

The internal energy \ $U$ \  is  the
 partial Legendre transformation  of the potential
\ $W (\rho_1 , \rho_2 , {\bf w})$ \ with
respect to the variable \ ${\bf w}$:
$$
U =  W (\rho_1 , \rho_2 , {\bf w}) \ - {\partial W\over \partial {\bf w}}
 \  {\bf w}
 = W + {\bf i}^* \ {\bf w}, \ \ \ \  {\bf i}^* =
 - {\partial W\over \partial {\bf w}},
$$
$$
W = U - {\bf i}^* {\bf w}, \ \ \ \
{\bf w}^*  =
{\partial U\over \partial {\bf i}}.
\eqno(2.3)
$$
The vector variable \ ${\bf i}$, which is also a Galilean invariant, can be called
{\it the relative  momentum}. Below, we will suppose that \ $U$ \ is invariant
under rotations (the case of {\it isotropic} media).
Hence,  \ $U$ \ is a
function of \ $i = \vert \ {\bf i} \ \vert$ \
and \ $W$ \ is a function of
\ $w = \vert \ {\bf w} \ \vert$.
In this case the relations (2.3) can be represented
in the form,
$$
 U = W - {\partial W\over \partial w}  \ w \ = \ W + i \ w, \ i \ =
\  -
 {\partial W\over \partial w}, \ W = U - i \ w, \ w \ =
\ {\partial U\over \partial i}.
\eqno(2.4)
$$

If \ $U$ \ does not depend on \ $i$, the variational principle (2.1), (2.2)
coincides with  classical Hamilton's
 principle of least  action: the Lagrangian is the difference of the kinetic
and the internal energy. The
 relation (2.3), (2.4) between \ $U$ \ and  \ $W$ \
will be justified later when the conservation law of  the total
 energy will be obtained.

\section{Governing Equations and Jump Conditions}

 Let \ $ {\bf x}$ \ be Eulerian coordinates,
\ ${\bf X}_\alpha$ be Lagrangian coordinates of the
\ $\alpha$-th  component, \ $\alpha\ = \ 1, 2$.
The relation between  Lagrangian and Eulerian coordinates is given by the
diffeomorphisms of the domain
\  ${\cal D}$ \ into  \ ${\cal D}$:
$$
{\bf x} = {\phiB}_{\alpha} \ (t , {\bf X}_{\alpha} ), \ {\bf X}_{\alpha}
={\psiB}_{\alpha} \ (t , {\bf x} ), \
{\phiB}_{\alpha} \ \circ \ {\psiB}_{\alpha} = I ,
\ \ \alpha = 1,2.
\eqno(3.1)
$$
Let
$$
F_{\alpha} = {\partial \ {\phiB}_{\alpha}\over \partial \  {\bf X}_{\alpha}},
 \ \alpha = 1,2
\eqno(3.2)
$$
be the deformation gradient at \ ${\bf X}_{\alpha}$ \ (or Jacobian matrix of the
mapping \ ${\phiB}_{\alpha}$). Let us  define the virtual
 motions of the mixture (Serrin [22],  Gouin [23] and others):
$$
{\bf x} = {\bf \Phi}_{\alpha} \ (t,\ {\bf X}_{\alpha}, \ \varepsilon_{\alpha}),
\ {\bf X}_{\alpha}\ = \ {\bf \Psi}_{\alpha} \ (t, \ {\bf x}, \
\varepsilon_{\alpha}),
 \  {\bf \Phi}_{\alpha} \ \circ \ {\bf \Psi}_{\alpha} \ = \ I,
\eqno(3.3)
$$
where \ $\varepsilon_{\alpha}$ \ varies in the neighbourhood of
zero. The real motion corresponds to \ $\varepsilon_{\alpha} = 0$:
$$
{\bf \Phi}_{\alpha} \ (t, \ {\bf X}_{\alpha}, 0) \ = \
{\phiB}_{\alpha} \ (t, \ {\bf X}_{\alpha}), \ {\bf \Psi}_{\alpha} \
( t, \ {\bf
x}, \ 0) \ = \ {\psiB}_{\alpha} \ ( t, \ {\bf x}).
$$
The associated variations \ $\delta_{\alpha} {\bf x}$ \ and \
$\delta {\bf X}_{\alpha}$ \ are defined by the relations:
$$
 \delta_{\alpha} {\bf x} \ = \
{\partial \ {\bf \Phi}_{\alpha}\over \partial \ \varepsilon_{\alpha}} \
 (t,{\bf X}_{\alpha}, 0) \ , \ \delta \ {\bf X}_{\alpha} \ = \
{\partial \ {\bf \Psi}_{\alpha}\over \partial \ \varepsilon_{\alpha}} \
 (t , {\bf x} , 0).
\eqno(3.4)
$$
The definitions (3.1) - (3.4) imply that
$$
\delta_{\alpha} {\bf x} \ = \ - F_{\alpha} < \delta \ {\bf X}_{\alpha} \ >.
\eqno(3.5)
$$
Let
$$
 f_{\alpha} (t,{\bf x}), \ \ \ \ \buildrel\circ\over f_{\alpha}\
(t, {\bf X}_{\alpha}) \ \equiv\  f_{\alpha} \ (t , {\phiB}_{\alpha}\
(t, {\bf X}_{\alpha}) \ ),
$$
$$
 \buildrel\land\over f_{\alpha}  \ (t,{\bf x}\ , \varepsilon_{\alpha} ), \ \ \ \
\buildrel\sim\over f_{\alpha} \ ( t, {\bf X}_{\alpha}, \varepsilon_{\alpha} )
\ \equiv \ \buildrel\land\over f_{\alpha} \ (t,{\bf \Phi}_{\alpha}
\ ( t, {\bf X}_{\alpha}, \ \varepsilon_{\alpha} ), \varepsilon_{\alpha}\  )
$$
be the unknown  quantities  of the \ $\alpha$-th  component (such as the density
 \ $\rho _{\alpha}$, the velocity \ ${\bf u}_{\alpha}$ \ etc.)  in Eulerian
and Lagrangian coordinates and their perturbations, respectively.
One defines Eulerian and Lagrangian variations
of the variable \ $f_{\alpha}$:
$$
 \delta \ f_{\alpha} \ = \ {\partial \buildrel\land\over
f_{\alpha}\over \partial \ \varepsilon_{\alpha}} \ (t , {\bf x} , 0) \ , \
\delta \ \buildrel\circ\over f_{\alpha}\ = \
\ {\partial \buildrel\sim\over f_{\alpha}\over
\partial\  \varepsilon_{\alpha}} \ (t , {\bf X}_{\alpha} , 0).
$$
It yields (Berdichevsky [17], Gouin [20,23])
$$
\delta \buildrel\circ\over f_{\alpha}\ = \ \delta \ f_{\alpha} \ + \ {\partial
f_{\alpha}\over \partial\  {\bf x}} \ \ \delta_{\alpha} {\bf x}.
\eqno(3.6)
$$
Using the Euler formula
$$
 \delta \ (det \ \buildrel\circ\over F_{\alpha} ) \ = \
det \ \buildrel\circ\over F_{\alpha} \ div (\delta_{\alpha}\ {\bf x}) \ , \
div (\delta_{\alpha}\ {\bf x}) \ = \ tr \bigg( {\partial\ \delta_{\alpha}\ {\bf x}
\over \delta \ {\bf x} }\bigg)
$$
and the mass balance (2.2) in the form
$$
\buildrel\circ\over \rho_{\alpha} \ det \
  \buildrel\circ\over F_{\alpha} \ = \buildrel\circ\over \rho_{\alpha}\
(0,{\bf X}_{\alpha}),
$$
we obtain Lagrangian variations of \ $\rho_{\alpha}$:
$$
\delta \ \buildrel\circ\over
\rho_{\alpha} \ =  -  \ \buildrel\circ\over \rho_{\alpha} \ \ div \ \delta_{\alpha}
 {\bf x}.
\eqno(3.7)
$$
We also note that
$$
\delta \ \buildrel\circ\over {\bf u}_{\alpha} \ = \ \ {\partial\over \partial \ t}
\ \ \delta_{\alpha}
 {\bf x}.
\eqno(3.8)
$$
Using  (3.6), (3.7), (3.8), we get

$$
 \delta \rho_{\alpha} \ = \ - \ div \ (\rho_{\alpha} \ \delta_{\alpha} {\bf x}) \
, \ \
\delta\ {\bf u}_{\alpha} \ = \ \ {d_{\alpha}\over dt}\
\delta_{\alpha} \ {\bf x}
\ - \
{\partial {\bf u}_{\alpha}\over \partial \ {\bf x}} \ < \ \delta_{\alpha} {\bf x}
\ >,
\eqno(3.9)
$$
$$
\ {d_{\alpha}\over dt}\ = \ {\partial\over \partial t} +  \
{\bf u}^*_{\alpha} \  \nabla^* ,
$$
where \ $\nabla^*$ \  means the gradient column operator.
We assume here that \ $\delta_{\alpha} {\bf x}$ \ are functions of Eulerian
variables. We define the vectors
$$
{\bf K}^*_{\alpha}  \ \equiv \
{1 \over \rho_{\alpha}} \,
{\partial L\over \partial {\bf u}_{\alpha}}
\ = \ {\bf u}^*_{\alpha} \ - \ (-1)^{\alpha} \
{1\over \rho_{\alpha}}
 \ \ {\partial W\over \partial w} \ {{\bf w}^*\over w},
\eqno(3.10)
$$
where
$$
L(\rho_1, \  \rho_2, \ {\bf u}_1, \ {\bf u}_2) = {1\over 2} \ \rho_1 \
\vert{\bf u}_1\vert^2 \ + \  {1\over 2} \ \rho_2 \ \vert{\bf u}_2\vert^2 \ - \ W
\ (\rho_1 , \rho_2 , w) \ , w\ = \vert\ {\bf u}_2- {\bf u}_1\vert.
$$
Varying  the variables \ $(\rho_1 ,{\bf u}_1)$ \ and \
$(\rho_2 ,{\bf u}_2)$ \  independently and denoting by \ $\delta_{\alpha}
{\cal I} $
the corresponding variation of the functional \ ${\cal I}$ \ , we obtain
$$
 \delta_{\alpha}{\cal I} = \int^{t_2}_{t_1}\int_{\cal D}\bigg( \
\delta\rho_{\alpha}\  \bigg ( {1\over 2} \vert {\bf u}_{\alpha}\vert^2 \ \ - \
{\partial
W\over \partial \rho_{\alpha}}\ \bigg) \ + \ \rho_{\alpha} \ {\bf K}_{\alpha}^* \
\delta \ {\bf u}_{\alpha}\ \bigg) \ d {\bf x}  \ dt.
$$
Taking now into account  the formulae  (3.9), we get
$$
\delta_{\alpha} {\cal I} = \int^{t_2}_{t_1} \ \int_{\cal D}\bigg( \ - div \
(\rho_{\alpha} \delta_{\alpha}{\bf x}) \
\bigg( \ {1\over 2 } \ \vert{\bf u}_{\alpha} \vert^2 \ -
\ {\partial W\over \partial
\rho_{\alpha}} \ \bigg) \ +
$$
$$
 + \ \rho_{\alpha} {\bf K}_{\alpha}^* \
\bigg( \ {d_{\alpha}\over dt } \ \delta_{\alpha} {\bf x} \ - \
{\partial {\bf u}_{\alpha}\over \partial {\bf x}} \ < \ \delta_{\alpha} {\bf x}
\ >  \ \bigg) \bigg) \ d {\bf x}  \ dt\ =
$$
$$
 = \ \int^{t_2}_{t_1} \ \int_{\cal D}
\ \Biggl\{ \ - \rho_{\alpha}\
\delta_{\alpha} {\bf x}^* \ \bigg( \ {\partial {\bf K}_{\alpha}\over \partial t
}\ + \ {\partial {\bf K}_{\alpha}\over \partial {\bf x}} \
< {\bf u}_{\alpha} \ >  \ + \ {\partial{\bf u}_{\alpha}^{*}\over \partial {\bf x}}
\ < \  {\bf K}_{\alpha} \ > \ +
$$
$$
+ \ \nabla^* \ \bigg({\partial W\over \partial
\rho_{\alpha}}  \
-  {\vert{\bf u}_{\alpha}\vert^2 \over 2}
 \ \biggr) \ \biggr) \
 + \ {\partial\over \partial t} \ (\rho_{\alpha} {\bf K}_{\alpha}^*
\delta_{\alpha}{\bf x}) \ +
$$
$$
+ \  div \ \bigg(\bigg( \rho_{\alpha}{\bf u}_{\alpha}
{\bf K}_{\alpha}^* \ + \  \rho_{\alpha}\ \bigg( {\partial W\over \partial
\rho_{\alpha}} \ - \ {\vert{\bf u}_{\alpha}\vert^2 \over 2}  \bigg) \ I \bigg)
 \ < \
\delta_{\alpha} {\bf x} \ > \ \bigg) \Biggr\} \ d{\bf x}  \ dt\  = 0.
\eqno(3.11)
$$

The mapping  \ $I$ \  means here  the unit tensor and, as  previously defined
in  the Introduction, for any vectors \ ${\bf a}$ \ and
\ $ {\bf b}$, \  ${\bf a}^* \ {\bf b}$ \  is  the scalar
product and
  \ ${\bf a}\ {\bf b}^*$ \  is  the tensor product. If all the functions in (3.11)
are smooth in the domain \ ${\cal D}$ \ and the variations \
$\delta_{\alpha}{\bf x}$ vanish \ on \ $\partial {\cal D}$, the divergence terms
 do not play any role. Therefore, we obtain the equations of motion:
$$
 {\partial{\bf K}_{\alpha}\over \partial t} \ + \
{\partial{\bf K}_{\alpha}\over \partial {\bf x}}\ <  {\bf u}_{\alpha} \ > \ + \
{\partial{\bf u}^*_{\alpha}\over \partial{\bf x}} \ < {\bf K}_{\alpha}\ > \ + \
\nabla^* \ \bigg({\partial W\over \partial
\rho_{\alpha}} \ -
\ {\vert{\bf u}_{\alpha}\vert^2 \over 2} \bigg) \ = \  0,
\eqno(3.12)
$$
or, equivalently,
$$
{\partial{\bf K}_{\alpha}\over \partial t }
\ + \ rot  \ {\bf K}_{\alpha}
\ \times \ {\bf u}_{\alpha} \ + \nabla^* \ \bigg({\partial W\over \partial
\rho_{\alpha}} \ -
\ {\vert{\bf u}_{\alpha}\vert^2 \over 2} \ +
{\bf K}_{\alpha}^* \ {\bf u}_{\alpha} \bigg) \ = \  0.
\eqno (3.12')
$$
If the function \ $W \  (\rho_1, \ \rho_2, \ w)$ \
 (or the internal energy \
$U  (\rho_1, \ \rho_2, \ i)$)
 is given, the equations (3.10), (3.12) with the mass conservation laws (2.2)
form the closed system of the governing
 equations.  We will show in Section 6 that, under natural restrictions on the
internal energy \ $U$ (or \ $W$),  the system
 is hyperbolic if the relative velocity is
sufficiently small.

 Now,  suppose that the domain \ $\displaystyle {\cal D} \times \lbrack \ t_1 ,
t_2 \ \rbrack $ \ is divided by a singular surface\ $S(t)$ \
having at any of its points
the normal unit vector \ ${\bf n }$ \ and the
 normal speed
 of displacement \ $D_n$.
Suppose also that at any point of \ $S (t)$ \ the right and left limits of \
 ${\bf K}_{\alpha}$ \ and \ $\rho_{\alpha}$ \ exist, but not necessary equal.
Then, the divergence terms in (3.11) give the jump conditions
(the Rankine-Hugoniot conditions):

$$
 \biggl[ {\bf n}^* \bigg( \rho_{\alpha }\ {\bf u}_{\alpha}\ {\bf K}_{\alpha}^* \
 + \  \rho_{\alpha }\  \bigg( \ {\partial W\over \partial\rho_{\alpha }} \ - \
{\vert{\bf u}_{\alpha}\vert^2 \over 2} \ \bigg) \ I \ \bigg)
\ <\delta_{\alpha} {\bf x} \ > \ -
$$
$$
- \  D_n \ \rho_{\alpha} \  {\bf K}_{\alpha}^*
\ \delta_{\alpha} {\bf x} \ \biggr]
\ = \  0 \ ,
\eqno(3.13)
$$
where the square brackets denote the jump.

If the singular surface \ $S (t)$ \ is a \ {\it shock wave} \ for \ $\alpha$-th
constituent (i.e.\ \ ${\bf n}^* \ {\bf u}_{\alpha} \ - \ D_n \ \not= \ 0$ \
 which means
 that the particles cross the surface), the formula (3.13) can be
symplified. Indeed, using, (3.5)
 and the fact that \ $\lbrack \ \delta  {\bf X}_{\alpha} \ \rbrack \
= \ 0$, we obtain:
$$
 \biggl[ \rho_{\alpha} \ (\ {\bf n}^* \ {\bf u}_{\alpha} \ - \ D_n)
{\bf K}_{\alpha}^* \ F_{\alpha} \ + \ \rho_{\alpha}\
\bigg({\partial W\over \partial\rho_{\alpha }} \ - \
{\vert{\bf u}_{\alpha}\vert^2 \over 2} \ \bigg) {\bf n}^* F_{\alpha}
\biggr]
\ = \  0.
\eqno(3.14)
$$

The equations (3.14) contain not only \ ${\bf K}_{\alpha}$ \ and \
$\rho_{\alpha}$ but also the deformation
gradient \ $F_{\alpha}$.  However, in the next section, we will obtain
shock conditions in
terms of  ${\bf K}_{\alpha}$ \ and \
$\rho_{\alpha}$,
 using a conservative form of the governing equations in Eulerian
coordinates.

Finally note that the governing equations could be  also obtained by using the
method of the Lagrange multipliers.

\section{Conservation Laws}

{\it Conservation laws},  i.e.\ the expressions of the form
$$
{\partial\ P_0\over \partial t}\ + \ div\ {\bf P}\ = \ 0,
$$
where \ $P_0$, \, ${\bf P}$ \ are functions of unknown variables, play an
important role in the theory of hyperbolic equations
(see, for instance, text-books by Serre [24]  or Smoller [25]).
The property of  conservativeness of  mathematical models, when
the number of linear independent
conservation laws admitted by the model is not less than the number of unknown
variables, is necessary to determine  weak solutions of the system.
Some conservation laws can play  a role of entropy, i.e.\  all
admissible solutions of the system of conservation
 laws must satisfy the "{\it entropy inequality}"
$$
 {\partial\ h_0 \over \partial\ t}\ + \ div \ {\bf h}\
\leq \ 0 \ ,
$$
where \ $h_0$,\, ${\bf h}$ \  are functions of the unknown quantities.

 The equations of motion (2.2), (3.12) admit two obvious additional  conservation
laws of the {\it  total
 momentum} and the {\it total
energy}, corresponding to the invariance of the Lagrangian
$$
 L = {1\over
2} \ \rho_1 \ \vert {\bf u}_1\vert^2 \ + \
{1\over 2} \ \rho_2  \ \vert {\bf u}_2 \vert^2 \ - \ W \ (\rho_1\ , \ \rho_2\ , w)
$$
 with respect to time and space shifts. They can be
obtained either using the
theorem of E.Noether (Olver [26], Ovsyannikov [27]) or
by direct calculations. The momentum conservation law is obtained  multiplying the
equations (3.12) by
 $\rho_{\alpha}$ \ and then  summing:
$$
{\partial\over \partial \ t} \ (  \ \rho_1 \ {\bf u}^*_1 \ +  \ \rho_2 \
{\bf u}^*_2)
 \ + \ div \ \bigg(  \ \rho_1 \ {\bf u}_1 \  {\bf u}^*_1 \ +
\rho_2 \  {\bf u}_2\ {\bf u}^*_2 \ -
$$
$$
- \
{\partial W\over \partial w} \ {{\bf w}\  {\bf w}^*\over w} \ + \
\bigg(\  \rho_1 \ {\partial W\over \partial \rho_1} \ + \ \rho_2
\ {\partial W\over \partial \rho_2} \ - \ W \bigg)\ I \ \bigg) =0.
\eqno(4.1)
$$
It admits also an alternative form in terms of \ ${\bf K}_{\alpha}$ \
(see the definition (3.10)):
$$
{\partial\over \partial \ t} \ (  \ \rho_1 \ {\bf K}^*_1 \ +  \ \rho_2 \
{\bf K}^*_2)  \ +
$$
$$
+ \ div \ \bigg(  \ \rho_1 \ {\bf u}_1 \  {\bf K}^*_1 \ +
\rho_2 \  {\bf u}_2\ {\bf K}^*_2 \ + \
\bigg(\  \rho_1 \ {\partial W\over \partial \rho_1} \ + \ \rho_2
\ {\partial W\over \partial \rho_2} \ - \ W \bigg)\ I \ \bigg) =0.
\eqno(4.1')
$$

Multiplying (3.12) by \ $\rho_{\alpha} \ {\bf u}_{\alpha}$ \ and then summing, we
obtain the energy conservation law
$$
{\partial\over \partial \ t} \
\bigg(\  \rho_1 {\vert {\bf
u}_1\vert^2\over 2} \ + \  \rho_2 {\vert {\bf u}_2\vert^2\over 2} \ + \
W \ - \ w \ {\partial W\over \partial w} \ \bigg)
\ + \ div \bigg(\ \rho_1 {\bf u}_1\bigg(\ {\vert {\bf u}_1\vert^2\over 2} \ + \
{\partial W\over \partial \rho_1} \ \ \bigg) \ +
$$
$$
+ \  \rho_2\ {\bf u}_1\ \bigg( \
{\vert {\bf u}_2\vert^2\over 2} \ + \  {\partial W\over \partial \rho_2} \ \bigg)
\  -  \ {\partial W\over \partial w} \
\bigg(\ {\bf u}_2 \ {\bf u}^*_2 \ - \ {\bf u}_1 \ {\bf u}^*_1 \ \bigg) \ < \ {{\bf
w}\over w} \ > \
\bigg) \ = \ 0.
\eqno(4.2)
$$
It admits an alternative form:
$$
{\partial\over \partial \ t} \
\bigg(\  \rho_1 {\vert {\bf
u}_1\vert^2\over 2} \ + \  \rho_2 {\vert {\bf u}_2\vert^2\over 2} \ + \
U \ \bigg) \ +
 \ div \  \bigg(\ \rho_1{\bf u}_1\ \bigg( \ {\partial W\over \partial \rho_1}\ - \
{\vert {\bf u}_1\vert^2\over 2} \ + \ {\bf K}^*_1{\bf u}_1\ \bigg) \ + \
$$
$$ + \rho_2 \ {\bf u}_2 \ \bigg( \ {\partial W\over \partial \rho_2 } \ - \
{\vert {\bf u}_2\vert^2\over 2} \ + \ {\bf K}^*_2{\bf u}_2\ \bigg)  \ \bigg)
= \ 0.
\eqno(4.2')
$$
The energy conservation laws (4.2), (4.2') explains now
the relation (2.4) between
\ $U$ \ and \ $\displaystyle W$:
$$
 \ U = W - w  \ {\partial W\over \partial w}.
$$

It follows from (3.12'), that an additional conservation law can be obtained, if
$$
rot \ {\bf K}_{\alpha}= \ 0.
\eqno(4.3)
$$
In that case
$$
{\partial {\bf K}_{\alpha}\over \partial \ t} \ + \ \nabla ^*\
\bigg( \ {\bf K}^*_{\alpha}\ {\bf u}_{\alpha} \ + \
 {\partial W\over \partial \rho_{\alpha} } \ - \
{\vert {\bf u}_{\alpha}\vert^2\over 2} \ \bigg)
\ = \  \ 0.
\eqno(4.4)
$$
The equations (4.3), (4.4)  are compatible. Indeed, the equation (4.4)  admits a
consequence
$$
{\partial\over \partial \ t} \
rot \ {\bf K}_{\alpha}= \ 0,
$$
and hence, (4.3)  can be considered as a restriction for the initial data.

To our knowledge, there is no additional conservation laws in terms of the
variables \ $\rho_{\alpha}$ \ and \ ${\bf u}_{\alpha}$.
 Finally the system  (2.2), (3.12)  contains eight desired
variables
 \ $\rho_{\alpha}$,\, ${\bf u}_{\alpha}$, \,  $\alpha = 1,2$.
If \  $ rot \ {\bf K}_{\alpha}\ \not= \ 0$, it  admits only six conservation
laws  (2.2), (4.1), (4.2). Hence, in the general
 case, the system seems not to be conservative.

Below, we will extend the set of desired variables, considering the deformation
gradients \ $F_{\alpha}$ \ as the
unknown quantities. We will show that the extended
system is a system of conservation laws.

Straightforward calculations show that \ $F_{\alpha}$ \ satisfies the equation:
$$
div \ \bigg( \ {F_{\alpha}\over det \ F_{\alpha}} \ \bigg) \ = \ 0.
\eqno (4.5)
$$
Using the Euler formula
$$
{d_{\alpha}\over d t} \ (\ det\  F_{\alpha}\ ) \ =
\  det\  F_{\alpha} \ div \ {\bf u}_{\alpha}, \ \ \ \ \
{d_{\alpha}\over d t} \
= \ {\partial\over \partial t} \ +  \ {\bf u}_{\alpha} \  \ \nabla \ ,
$$
we obtain
$${d_{\alpha}\over d t} \ \bigg( \ {F_{\alpha}\over det \ F_{\alpha}} \ \bigg) \ =
\  \bigg( \ {\partial{\bf u}_{\alpha}\over \partial {\bf x}} \ -
 \  div \ {\bf u}_{\alpha} \ I \ \bigg) \ {F_{\alpha}\over det \ F_{\alpha}}.
\eqno(4.6)
$$
Taking into account  (4.5), we rewrite (4.6)  in the divergence form
$$
{\partial\over \partial t} \
\bigg( \ {(F_{\alpha})^m_s \over det \ F_{\alpha}} \ \bigg) \ + \
{\partial\over \partial x^k} \
\bigg( \ {({\bf u}_{\alpha})^k \ (F_{\alpha})^m_s \ - ({\bf u}_{\alpha})^m \
(F_{\alpha})^k_s\over det \ F_{\alpha}}  \ \bigg)
= \ 0.
\eqno(4.7)
$$
Here $\ (F_{\alpha})^m_s \ $ are the components of \ $F_{\alpha}$ \
($m$ denotes the lines and $s$
denotes the columns), \
$({\bf u}_{\alpha})^k$ \ are the components of
\ ${\bf u}_{\alpha}$.
The repeated latin indices imply summation. Conversely,
applying the operator \ $div$ \ to the equation (4.7), we obtain:
$$
{\partial\over \partial t} \ div\
\bigg( \ {F_{\alpha}\over det \ F_{\alpha}} \ \bigg) \ = \ 0.
$$
Hence, we can replace the equation (4.5)  by the evolution equation (4.7),
considering (4.5)  as the
 restriction for initial data: if the condition  (4.5)
is fullfilled at \ $t = 0$, it is valid for any
 time. We note that  the divergence form (4.7)  was earlier obtained
by Godunov and Romensky [28]  in the theory of elasticity.

Finally, by using
 (4.7), we  get the equation (3.12) in the conservative form:
$$
{\partial\over \partial t} \
\bigg( \
{{\bf K}^*_{\alpha}\ F_{\alpha}\over det  \ F_{\alpha} } \bigg)
\ + \ div \ \bigg( \ \bigg( \ {\bf u}_{\alpha}
\ {\bf K}^*_{\alpha}\ + \ \bigg( \ {\partial W\over \partial \rho_{\alpha}} \ - \
{\vert{\bf u}_{\alpha}\vert^2\over 2} \ \bigg) \ I \ \biggr) \
{F_{\alpha}\over det \ F_{\alpha}} \ \bigg) \ = \ 0,
\eqno(4.8)
$$
which represents the conservation of {\it local momentum} of $\alpha$-th
phase.
 In the next section, we
 will show that the conservation law (4.8)  corresponds to the jump
conditions (3.13) obtained from the variational
principle  (2.1).

The equations (2.2), (4.5), (4.7), (4.8)  are in the conservative form
relative to the
variables \ $\rho_{\alpha}$, \ ${\bf u}_{\alpha}$ \ and \ $F_{\alpha}$.
It is also shown that the equations admit  also conservation of
total momentum, conservation of
total energy and , in the case of \ $rot \ {\bf K}_{\alpha}
\ = \ 0$ , conservation of \ ${\bf K}_{\alpha}$.

\section{Analysis of Rankine-Hugoniot Conditions}

Let \ $S(t)$ \ be a singular surface with the unit normal vector
\ ${\bf n}$ \ and
the normal velocity \ $D_n$, where
 the functions \ $\rho_{\alpha}$, \  ${\bf K}_{\alpha}$ \ and \ $F_{\alpha}$ \
have jumps. The equations (2.2), (4.7), (4.8)  imply the following
 Rankine-Hugionot
conditions:
$$
\lbrack \ \rho_{\alpha}  \ ( \ {\bf n}^*\ {\bf u}_{\alpha} - D_n
\ ) \ \rbrack \ = \
0\ ,
\eqno(5.1)
$$
$$
 \lbrack \
 ( \ {\bf n}^* \ {\bf u}_{\alpha} - D_n \ ) \ {F_{\alpha}\over det \
F_{\alpha}} \ - \ {{\bf u}_{\alpha}\  \ {\bf n}^* \ F_{\alpha}\over det \
F_{\alpha}} \ \rbrack \ = \ 0 \ ,
\eqno(5.2)
$$
$$
\bigg[( \ {\bf n}^* \ {\bf u}_{\alpha} - D_n \ ) \ {\bf K}^*_{\alpha}
\ {F_{\alpha}\over det \
F_{\alpha}} \ + \ \bigg( \ {\partial W\over \partial \rho_{\alpha}} \ - \
{\vert{\bf u}_{\alpha}\vert^2\over 2} \ \bigg) \ {\bf n}^* \
{F_{\alpha}\over det \ F_{\alpha}} \bigg]
 \ = \ 0.
\eqno(5.3)
$$

 If \ ${\bf n}^* \ {\bf u}_{\alpha} - D_n \ = \ 0$, we call it a
{\it  contact  discontinuity}.
If \ ${\bf n}^* \ {\bf u}_{\alpha} - D_n \ \not= \ 0$,
 we call it a   {\it  shock  wave}.

Let us consider the case of contact discontinuity. It follows from (5.2) - (5.3)
that
$$
\bigg[ \ {{\bf u}_{\alpha}  {\bf n}^* \
F_{\alpha}\over det \ F_{\alpha}} \bigg]
 \ = \ 0,
\eqno (5.2')
$$
$$
\bigg[ \ \bigg( \ {\partial W\over \partial \rho_{\alpha}} \ - \
{\vert{\bf u}_{\alpha}\vert^2\over 2} \ \bigg) \ {{\bf n}^* \
F_{\alpha}\over det \ F_{\alpha}} \  \bigg] \ = \ 0.
\eqno(5.3')
$$
Multiplying (5.2')  by the normal vector \ ${\bf n}^*$ \ and taking into account
that \ $\lbrack \ {\bf n}^*{\bf u}_{\alpha} \ \rbrack \ = \ 0$, we get
$$
\bigg[ \ {{\bf n}^* \ F_{\alpha}\over det \ F_{\alpha}} \ \bigg] \ = \  0.
\eqno (5.2'')
$$
The equations  (5.2')  and  (5.2'')  imply that the  velocity
 ${\bf u}_{\alpha}$ \ is continuous.
Hence, it follows from (5.2'), (5.3')  that the conditions
on the contact discontinuity are given by
$$
\lbrack \ {\bf u}_{\alpha} \ \rbrack \ = \  0, \ \
\bigg[ \ {\partial W\over \partial \rho_{\alpha}} \
 \bigg] \ = \ 0.
\eqno(5.4)
$$

Let us consider shock waves. First of all, we note that the condition  (5.1) can be
rewritten in the form
$$
\lbrack \ \rho_{\alpha} \ det \ F_{\alpha} \ \rbrack \ = \  0.
\eqno(5.1')
$$
Hence, the relations  (3.14)  derived from the variational principle coincide with
(5.3). As in the case of the
 contact discontinuity, we rewrite  the equations (5.1) -
(5.3)   in terms of ${\bf u_{\alpha}}$, \, $\rho_{\alpha}$.
 Multiplying  (5.2)  by \
${\bf n}^* $, we get the equation (5.2''). Further, let  $\ S(t)\ $ be a  shock
surface  in Eulerian coordinates, and \  $S_{\alpha} (t)$\  be its image
 in Lagrangian coordinates of the \ $\alpha - th$ \ component.
Let \ ${\bf q}_{\alpha}$ \ be a tangent vector to  \ $S_{\alpha} (t)$.
Then, \  ${\bf q} \ = \ F_{\alpha} \, < {\bf q}_{\alpha} >$ \
is a tangent  vector to  \ $S(t)$.
  Multiplying (5.3)  from the right by \
${\bf q}_{\alpha}$ \ and taking into account that \
${\bf n}^* \ {\bf q} \ = \ 0$,
 we obtain:
$$
\bigg[ \ ({\bf n}^* \ {\bf u}_{\alpha} - D_n) \
{{\bf K}^*_{\alpha} \ {\bf q}\over det \ F_{\alpha}} \ \bigg] \ = \ 0.
$$
It follows then from the last relation and from equations  (5.1),  (5.1')
that the tangential
component \ ${\bf K}_{\alpha q}$ \ of
\ ${\bf K}_{\alpha} \ =
\ {\bf n} \ ( {\bf n}^* \ {\bf K}_{\alpha}) \ + \ {\bf K}_{\alpha q}$,
\  ${\bf n}^* \ {\bf K}_{\alpha q}\ = \ 0$,  is continuous:
$$
\lbrack \ {\bf K}_{\alpha q}\ \rbrack = \ 0.
\eqno (5.5)
$$
Multiplying (5.2)  by  \ ${\bf K}^*_{\alpha q}$, we have:
$$
\bigg[ \ {\bf K}^*_{\alpha q} \ F_{\alpha} \ \
{({\bf n}^* \ {\bf u}_{\alpha} \ - D_n)\over  det \ F_{\alpha} } \ \bigg] \ = \
\bigg[ \ {({\bf K}^*_{\alpha q}{\bf u}_{\alpha} ) \ {\bf n}^* \ F_{\alpha}\over
det \ F_{\alpha} } \ \bigg].
\eqno(5.6)
$$
Replacing  in (5.3)  \ ${\bf K}_{\alpha}$ \  by
$\ {\bf K}_{\alpha q}\ + {\bf n}\ ({\bf n}^* \ {\bf K}_{\alpha})\ $, we get
$$
\bigg[ \ ({\bf n}^* \ {\bf u}_{\alpha} \ - D_n) \
{({\bf K}^*_{\alpha q}+ {\bf n}^* \ ({\bf n}^*\ {\bf K}_{\alpha})\ ) \
F_{\alpha}\over  det \ F_{\alpha}} \ +
$$
$$
+ \
\bigg( \ {\partial W\over \partial \rho_{\alpha}} \ - \
{\vert \ {\bf u}_{\alpha}\ \vert^2\over 2 } \ \bigg) \
{{\bf n}^* \ F_{\alpha}\over  det \ F_{\alpha}} \ \bigg] \ = \ 0.
\eqno(5.7)
$$
Equations  (5.2''), (5.6), (5.7) imply then that
$$
\bigg[ \ {\partial W\over \partial \rho_{\alpha}} \ - \
{\vert \ {\bf u}_{\alpha}\ \vert^2\over 2 } \ + {\bf K}^*_{\alpha}\
{\bf u}_{\alpha} \ - D_n \ {\bf K}^*_{\alpha}{\bf n} \  \bigg] \ =
 \  0.
\eqno (5.8)
$$
Equations  (5.1),  (5.5), (5.8) are the Rankine-Hugionot conditions for shocks in
terms of the  variables ${\bf u_{\alpha}, \rho_{\alpha}}$.
It is worth to note that the jump conditions (5.5), (5.8)
 for shocks coincide with the
jump conditions for the
equation (4.4). Nevertheless, we did not use in our
derivation the hypothesis \  $rot\ {\bf K}_{\alpha} \ = 0$.

The conservation laws (4.1), (4.2) imply also additional jump
conditions for the total momentum and energy. Hence, we obtain
an {\it overdetermined} system of the Rankine-Hugoniot conditions
in terms of the physical variables \ ${\rho}_{\alpha}$,\, ${\bf u}_{\alpha}$,\,
${\alpha} \, = \, 1, 2$. This is a consequence of the well-known fact that
the same system of equations can be written in different divergence forms each of
which defines different weak solution (see, e.g., [24], [25]). We can now
question, which divergence form is the more appropriate? For one-velocity
systems this choice is unambiguous. For example, for isentropic gas flows we
use the conservation of mass and momentum. The mechanical energy plays the role
of entropy:  it decreases through the shocks [24].
The choice of appropriate shock conditions for the two-velocity case is less
clear. Hamilton's principle provides
a set of Rankine-Hugoniot conditions (3.14).
As was shown in Section 4, these last conditions correspond to the divergence
form
(4.8)  which represents the conservation of local momentum. Formally,
equations (5.5), (5.8), which are issued from (3.14) and
supplemented by the equations of mass conservation (5.1), form
a complete set of   Rankine-Hugoniot conditions.
Similar to the one-velocity isentropic case, the energy conservation law
(4.2) (or (4.2')) should apparently play for shocks the role of "entropy"
inequality:
$$
- D_n \ \lbrack \ E \ \rbrack \ + \
\bigg[ \ \sum^2_{\alpha=1} \ \rho_{\alpha} \ {\bf n}^*\ {\bf u}_{\alpha}
\ \bigg( {\partial W\over \partial \rho_{\alpha}} \ - \
{\vert \ {\bf u}_{\alpha}\ \vert^2\over 2 } \ + {\bf K}^*_{\alpha}\
{\bf u}_{\alpha} \  \bigg) \ \bigg] \ \leq \ 0,
$$
$$
E \ =  \  \rho_1 {\vert {\bf
u}_1\vert^2\over 2} \ + \  \rho_2 {\vert {\bf u}_2\vert^2\over 2} \ + \ U.
$$
The jump conditions obtained are inconsistent with the conservation of
the total momentum (4.1) (or (4.1')):
$$
\bigg[ \sum^2_{{\alpha}=1} \ \rho_{\alpha} ({\bf n}^* \ {\bf u}_{\alpha}- D_n )
 \ {\bf K}_{\alpha}^* + {\bf n}^* \ \bigg( \ \rho_1 \ {\partial W\over
\partial\rho_1} \ + \  \rho_2 \ {\partial W\over \partial\rho_2} \ - \ W
\bigg)  \ \bigg] \ = \ 0.
$$
Finally, we note that the system of the jump conditions
for two-fluid  models is generally underdetermined. This does not permit
to define weak solutions. In our case, this system is overdetermined.
Hamilton's principle provides a complete set of appropriate
jump conditions. Is that choice  correct?
Only physical arguments are able to give a definite answer to this
question.

\section{Hyperbolicity}

The property of hyperbolicity of governing equations is very important,
because it
 implies the well-posedness of the Cauchy problem. Below, we will give
a sufficient condition
 of the hyperbolicity of the system (2.2), (3.12) in the
multi-dimensional case provided that \ $ rot \ {\bf K}_{\alpha} \ = \ 0$.

First, we transform our system to  a symmetric form. Considering  the
Lagrangian of our system
$$ L = \sum^2_{\alpha=1} \ {1\over 2} \
\rho_{\alpha} \ \vert{\bf u}_{\alpha}  \vert^2 - W \ (\rho_1\ , \rho_2\ , w),
$$
we get:
$$
 d L = \sum^2_{\alpha=1} \ \bigg(\ {\partial L\over \partial
\rho_{\alpha}} \ d \rho_{\alpha} \ + \ {\partial L\over \partial {\bf u}_{\alpha}}\
\ d {\bf u}_{\alpha} \ \bigg) \ = \ \sum^2_{\alpha=1} \ \bigg(
{\partial L\over \partial
\rho_{\alpha}} \ d \rho_{\alpha} \ + \ {1\over \rho_{\alpha}} \  {\partial L\over
\partial {\bf u}_{\alpha}} \ \rho_{\alpha} \ d {\bf u}_{\alpha} \ \bigg) \ =
$$
$$
= \sum^2_{\alpha=1} \ \bigg(
{\partial L\over \partial\rho_{\alpha}} d \rho_{\alpha} \  + \ {\bf K}^*_{\alpha} \
( d \  (\rho_{\alpha} \ {\bf u}_{\alpha}) -   {\bf u}_{\alpha}
d \ \rho_{\alpha} \ ) \ \bigg) \ =
$$
$$
 = \sum^2_{\alpha=1} \ \bigg(\ \bigg(
{\partial L\over \partial\rho_{\alpha}} \ - {\bf K}^*_{\alpha} \ {\bf u}_{\alpha}
\ \bigg)  \ d \ \rho_{\alpha}\ + \ {\bf K}^*_{\alpha}
\ d (\rho_{\alpha} \ {\bf u}_{\alpha})\ \bigg)  = \
\sum^2_{\alpha=1} \ ( \sigma_{\alpha} \ d \rho_{\alpha} + {\bf K}^*_{\alpha} \ d
{\bf j}_{\alpha} \ ) \ ,
$$
where
$$\sigma_{\alpha} \ = {\partial L\over \partial \rho_{\alpha}} \ - \
\ {\bf K}^*_{\alpha} \ {\bf u}_{\alpha} \ = \ - \
\bigg(
{\partial W\over \partial \rho_{\alpha}}\ - {1\over 2} \ \vert{\bf u}_{\alpha}
\vert^2\ + \ {\bf K}^*_{\alpha} \ {\bf u}_{\alpha} \ \bigg) \ , \
{\bf j}_{\alpha} \ =\ \rho_{\alpha}{\bf u}_{\alpha}.
$$
Or, equivalently,
$$
 d\ (L - \ \sum^2_{\alpha=1}\ \sigma_{\alpha}\
\rho_{\alpha} )\  = \ \sum^2_{\alpha=1} \ - \ \rho_{\alpha}d \sigma_{\alpha} \ + \
 {\bf K}^*_{\alpha} \ d
{\bf j}_{\alpha}.
\eqno(6.1)
$$
Let us introduce
$$
G (\sigma_{1},\ \sigma_{2}, \ {\bf j}_{1}, \ {\bf j}_{2}) \ = \ L (\rho_{1}, \
\rho_{2},\ {\bf j}_{1}, \ {\bf j}_{2}) \ - \  \sum^2_{\alpha=1}\ \sigma_{\alpha}\
\rho_{\alpha} \ = \ L\ - \ \sum^2_{\alpha=1}\
 {\partial L\over \partial\rho_{\alpha}} \ \rho_{\alpha}.
\eqno(6.2)
$$
The function \ $G$ \ is a partial Legendre transformation of
\ $L\ (\rho_{1}, \  \rho_{2},\ {\bf j}_{1}, \ {\bf j}_{2})$ \
with respect to the variables
\ $\rho_{\alpha}$:
$$
{\partial G\over \partial \sigma_{\alpha}} \ = \ - \
\rho_{\alpha}, \ \ \ \  {\partial G\over \partial {\bf j}_{\alpha}} \ = \ {\bf
K}^*_{\alpha}.
\eqno(6.3)
$$
By using (6.1) - (6.3)  we get:
$$
{\partial \over \partial t} \bigg( \ {\partial G\over \partial
\sigma_{\alpha}} \ \bigg) \ - \ div \ {\bf j}_{\alpha} \ = \ 0,
$$
$$
{\partial \over \partial t}\bigg( \ {\partial G\over \partial {\bf
j}_{\alpha}}  \ \bigg)
\ - \ \nabla \ \sigma_{\alpha} \ = \ 0.
$$

Or
$$ {\partial \over \partial t}\bigg( \ {\partial G\over \partial
\sigma_{\alpha}} \ \bigg) \ - \ div \ \bigg( \ {\partial \over \partial
\sigma_{\alpha}}\ \bigg( \ \sum^2_{\beta=1} \ \sigma_\beta  \ {\bf j}_{\beta} \
\bigg)\ \bigg)
 \ = \ 0,
\eqno(6.4)
$$
$$
{\partial \over \partial t}\bigg( \ {\partial G\over \partial {\bf
j}_{\alpha}} \ \bigg) \ - \ div \ \bigg( \ {\partial \over \partial {\bf
j}_{\alpha}}\ \bigg( \ \sum^2_{\beta=1} \ \sigma_{\beta } \ {\bf j}_{\beta} \
\bigg)\ \bigg)
 \ = \ 0.
\eqno(6.5)
$$

The system (6.4), (6.5) can be rewritten in a symmetric form  \
(Friedrichs [29],  Friedrichs \& Lax [30],  Godunov [31],  Godunov \&  Romensky
[28]):
$$
A  \ {\partial {\bf u}\over \partial t} \ + \ B^i \
{\partial {\bf u}\over \partial x^i} \ = \ 0 \ \ , \ \ A = A^* \ , B^i \ = (B^i)^*,
\ \ \  i = 1,\, 2,\, 3,
\eqno(6.6)
$$
where
$$
{\bf u}^* \ = \ (\sigma_1 \ ,\  \sigma_2 \ ,
{\bf j}^*_1,\ {\bf
j}^*_2) \ , \hskip 0,5 cm
A \ = {\partial^2 G\over \partial {\bf u}^2}
$$
and the matrices \ $B^i$ \ can be  obtained from  (6.4), (6.5).
If, moreover, the matrix \ $A$ \ is positive, the system (6.6) is hyperbolic.
It is
worth to note that equations
(6.4), (6.5) admit a conservation law of the form
$$
  {\partial \over \partial t}\bigg( \ \sum^2_{\alpha=1} \
\sigma_{\alpha}  \  {\partial
G\over \partial \sigma_{\alpha}} \ + \ {\partial G\over \partial {\bf
j}_{\alpha}}
 \ {\bf j}_{\alpha} - G \bigg) \ - \ div \
\bigg( \ \sum^2_{\alpha=1} \ \sigma_{\alpha}  \ {\bf j}_{\alpha} \bigg) \
= \ 0,
$$
which coincides with the equation of energy (4.2').
(see also Godunov [31]).

So, we need to prove  convexity of \ $G \ (\sigma_{1},\ \sigma_{2},\
{\bf j}_{1}, \ {\bf j}_{2})$.
 As we have previously mentioned, \ $G$ \ is the Legendre transformation
 of \ $ L \
(\rho_1, \ \rho_2, \ {\bf j}_1, \ {\bf j}_2 \ )$ \ with respect to \
$\rho_1$, \, $\rho_2$ (see the formulae (6.2) \ , (6.3)).
If \ $L$ \ is a
 {\it  convex}  function  \ with respect to \ ${\bf j}_1$, \,  ${\bf
j}_2$ \ and  \ {\it concave} \ with respect to \ $\rho_1$, \,  $\rho_2$,
 then
\ $G  \ (\sigma_1,\  \sigma_2, {\bf j}_1, \ {\bf j}_2 \ )$  will be
convex, that means  the hyperbolicity of our system.
 Hence, it is sufficient to prove
that the symmetric matrices \ $L _{{\bf j}{\bf j}}$ and \ $L_{\rho\rho}$ ,
defined below, are positive and negative
 definite, respectively:
$$
L _{{\bf j}{\bf j}} \ \equiv \pmatrix{L _{{\bf j}_1{\bf j}_1} &
L _{{\bf j}_1{\bf j}_2}\cr\cr
L _{{\bf j}_1{\bf j}_2} & L _{{\bf j}_2{\bf j}_2}} \ > \ 0,
$$
$$
L_{\rho\rho} \ \equiv \
\pmatrix{L _{{\rho}_1{\rho}_1} & L _{{\rho}_1{\rho}_2} \cr\cr
 L _{{\rho}_1{\rho}_2}  & L _{{\rho}_1{\rho}_2} }\ \  < \ \ 0.
$$
It follows from (1.2) that for the isotropic case
$$
L \ (\rho_1, \rho_2, \ {\bf j}_1, {\bf j}_2) \ = \ {\vert
{\bf j}_1\vert ^2 \over  2\rho_1} + {\vert {\bf j}_2\vert ^2 \over 2\rho_2} - W
\bigg(\rho_1, \rho_2, \left\vert{{\bf j}_2\over \rho_2}\ -{{\bf
j}_1\over \rho_1}\right\vert\ \bigg).
$$
Then,
$$
L_{\bf j_{\alpha}} ={{\bf
j_{\alpha}}^{*}\over \rho_{\alpha}}  \ - \ (-1)^{\alpha} {1\over \rho_{\alpha}}\
{\partial W\over \partial w}\
{{\bf w}^{*}\over w},
$$
$$
L_{{\bf j}_1 {\bf j}_2} ={1\over \rho_1 \rho_2}\
{\partial^2 W\over \partial
w^2}\ {{\bf w}{\bf w^*}\over w^2},
$$
$$
L_{{\bf j}_{\alpha} {\bf j}_{\alpha}} \  = \
{1\over \rho_{\alpha} }
I -
{1\over \rho^2_{\alpha}}
{\partial^2 W\over \partial w^2}
 {{\bf w}{\bf w}^*\over w^2}.
$$
Straightforward  calculations shows that $\ L_{\bf j \bf j}\ > \ 0$,
if \ $\displaystyle {\partial^2
W \over \partial w^2} \ < 0$.

Now, we calculate \ $\displaystyle  L_{\rho\rho}$ :
$$
{\partial L\over \partial \rho_{\alpha} }= \ - \
{\vert {\bf j}_{\alpha}\vert
^2 \over  2{\rho_{\alpha}}^2}\ - \ {\partial W\over \partial \rho_{\alpha}}\
+ \ (-1)^{\alpha}
{\partial W\over \partial w}\ {{\bf w^*} {\bf j}_{\alpha}\over w \rho_{\alpha}^2},
$$
$$
{\partial^2 L\over \partial\rho_1\partial\rho_2} \ = \
- \ {\partial^2 W\over \partial\rho_1\partial\rho_2}\ - \ {\partial\over
\partial\rho_2}\bigg(\ {\partial W\over \partial w} \ {{\bf w}^* {\bf j}_1\over
w\rho^2_1}\ \bigg),
$$
$$
{\partial^2 L\over \partial\rho^2_{\alpha}} \ = \
{\vert {\bf j}_{\alpha} \vert^2\over
\rho^3_{\alpha}}
\ - \ {\partial^2 W\over \partial\rho^2_{\alpha}} \ - \
{\partial\over \partial\rho_{\alpha}}
 \bigg( \ {\partial W\over \partial w}\
{ {\bf w}^*{\bf j}_{\alpha}\over w \rho^2_{\alpha} }
\bigg).
$$

Consequently, \ $L_{\rho\rho}\, <\, 0$\  if the velocities  \ ${\bf u}_\alpha$\
 are sufficiently
small, and the function \ $W$ \ is convex
with respect to \ $\rho_1$,\, $\rho_2$.  However, the governing equations  are
invariant under the Galilean group of transformations
$$
{\bf x}'={\bf x}+{\bf U} t, \ {\bf
u}'_{\alpha} = {\bf u}_{\alpha}  + {\bf U},\ t'= \ t.
$$
  That means that the condition "the velocities \ ${\bf u}_{\alpha}$\
 are
sufficiently small" can be replaced by "the relative velocity \ ${\bf w}$ \  is
sufficiently small".  Hence, the conditions
$$
{\partial^2 W\over \partial w^2}< \ 0 \ , \ {\partial^2
W\over\partial\rho_1^2}\ > \ 0, \ {\partial^2 W\over\partial\rho_1^2}\ {\partial^2
W\over\partial\rho_2^2}-\bigg( {\partial^2 W\over
\partial\rho_1\partial\rho_2} \bigg)^2 \ >0
\eqno(6.7)
$$
guarantee the hyperbolicity of our system for small relative velocity of phases.
Due to (2.4), the
inequalities (6.7) mean the  convexity  of the internal energy \
$\displaystyle U (\rho_1,\rho_2, i)$,   that
corresponds  to a natural condition of {\it thermodynamic  stability}.

Finally,  we have established that the thermodynamic  stability  implies  the
hyperbolicity  of the governing equations for small relative
 velocity \ ${\bf w}$,  provided that \
$rot \ {\bf K}_i = 0$.  The last condition   is always fullfilled for
 one-dimensional flows.

\acknowledgements{The authors thank D. Serre for helpful discussions.}


\begin{thebibliography}{99}
\bibitem{Ishii75}
Ishii, M., {\it Thermo-Fluid Dynamic Theory of Two-Phase Flow.} Paris,
Eyrolles, (1975).
\bibitem{Nigmatulin91}
Nigmatulin, R.I. {\it Dynamics of Multiphase Media}.
{\bf 1,2}. Hemisphere Publishing Corporation, New York, (1991).
\bibitem{Drew83}
Drew, D.A.
'Mathematical modeling of two-phase flow', {\it Annual  Rev.
Fluid Mech.} {\bf 15}  (1983), 261-291.
\bibitem{Kraiko65}
Kraiko, A.N. and Sternin, L.E., 'To the theory of flow of two-velocity
continuum with solid and liquid particles', {\it Prikl. Mat. i Mekh.} {\bf 29}
 (1965), 418-429.
\bibitem{Stuhmiller77}
Stuhmiller, J.H., 'The influence of  interfacial pressure forces on the
character of two-phase model equations', {\it Int. J. Multiphase Flow} {\bf 3}
(1977), 551-560.
\bibitem{Liapidevskii87}
Liapidevskii, V.Yu., 'Hyperbolic two-phase flow models based on
conservation laws',  {\it XI Intern. Symp. Nonlinear Acoustics}, part I,
 Novosibirk, (1987).
\bibitem{Ransom84}
Ransom, V.H. and Hicks, D.L., 'Hyperbolic two-pressure models for
two-phase flow',  {\it J. Comput. Physics} {\bf 53} (1984), 124-151.
\bibitem{Biesheuvel84}
Biesheuvel, A. \& Van Wijngaarden, L., 'Two-phase flow equations for
a dilute dispersion of gas bubbles in liquid', {\it J. Fluid Mech.} {\bf 148}
(1984), 301-318.
\bibitem{Fabre95}
Fabre, J., Lin{\' e}, A. and Gadoin, E., 'Void and pressure waves in
slug flow', {\it Proceedings of the IUTAM Symposium on  Waves in Liquid/Gas
 and  Liquid/Vapour Two-Phase Systems}, Kyoto, Japan, (1995).
\bibitem{Khalatnikov71}
Khalatnikov, I.M., {\it Theory of Superfluidity.} Moscow, Nauka, (1971).
\bibitem{Landau89}
Landau, L.D. and Lifshits, E.M., {\it Fluid Mechanics} Pergamon Press,
(1989).
\bibitem{Putterman74}
Putterman, S.,  {\it Super Fluid Hydrodynamics}, New York,
American Elsevier Publishing Company, (1974).
\bibitem{Dorovsky92}
Dorovsky, V.N. and  Perepechko, Yu.V., 'Phenomenological description of
two-velocity media with relaxing  shear stresses', {\it Prikl. Mekh.
Tekh. Fiz.} {\bf 3} (1992), 99-110.
\bibitem{Roberts87}
Roberts, P.H. and Loper, D.E., 'Dynamical processes in slurries', In:
{\it Structure and Dynamics of Partially Solidified System}. NATO ASI, serie E,
{\bf 125} (1987), 229-290.
\bibitem{Shugrin94}
Shugrin, S.M., 'Two-velocity hydrodynamics and
thermodynamics',
{\it Prikl. Mekh. Tekh. Fiz.} {\bf 4} (1994), 41-59.
\bibitem{Bedford78}
Bedford, A. and Drumheller, D.S.,'A variational
 theory of immiscible mixtures',
{\it Arch. Rat. Mech. Anal.} {\bf 68} (1978), 37-51.
\bibitem{Berdichevsky83}
Berdichevsky, V.L., {\it Variational Principles of Continuum
Mechanics}, Moscow, Nauka, 1983.
\bibitem{Geurst86}
Geurst, J.A., 'Variational principles and two-fluid hydrodynamics of
bubbly liquid/gas mixtures',  {\it Physica A}  {\bf 135} (1986), 455-486.
\bibitem{Geurst85}
Geurst, J.A., 'Virtual mass in two-phase bubbly flow',  {\it Physica A}
{\bf 129} (1985), 233-261.
\bibitem{Gouin90}
Gouin, H., 'Variational theory of mixtures in continuum mechanics',
{\it Eur. J. Mech}, B/Fluids {\bf 9} (1990), 469-491.
\bibitem{Gavrilyuk97}
Gavrilyuk, S.L., Gouin, H. and Perepechko, Yu.V., 'A variational  principle
for two-flud models', {\it C.R.
Acad. Sci. Paris}  S{\'e}rie II b, {\bf 324}  (1997),  p.483-490.
\bibitem{Serrin59}
Serrin, J.,  'Mathematical principles of  classical fluid mechanics',
{\it Encyclopedia of Physics}, {\bf VIII/1}, Springer Verlag (1959),
125-263.
\bibitem{Gouin76}
Gouin, H., 'Noether theorem in fluid mechanics', {\it Mech. Res.
Commun}, {\bf 36} (1976), 151-155.
\bibitem{Serre96}
Serre, D., {\it Syst\`emes de Lois de Conservation}. {\bf I,\ II} Diderot
Editeur, Arts et Sciences, (1996).
\bibitem{Smoller94}
Smoller, J.,  {\it Shock Waves and Reaction-Diffusion
Equations}. Springer Verlag, (1994).
\bibitem{Olver86}
Olver, P.J.,  {\it Applications of Lie Groups to Differential
Equation}.  Springer-Verlag, (1986).
\bibitem{Ovsyannikov78}
Ovsyannikov, L.V.,  {\it Group Analysis of Differential
Equations}.  Moscow, Nauka, (1978).
\bibitem{Godunov95}
Godunov, S.K. and Romensky, E.I., 'Thermodynamics, conservation laws
and symmetric forms of differential equations in mechanics of continuous media'.
In: {\it Computational Fluid Dymamics Review}. Ed. by M. Hafez, K. Oshima, John
Willey \& Sons, (1995).
\bibitem{Friedrichs58}
Friedrichs, K.O., 'Symmetric positive linear differential equations',
{\it Commun. Pure Appl. Math.} {\bf 11} (1958), 333-418.
\bibitem{Friedrichs71}
Friedrichs, K.O. and Lax, P.D., 'Systems of conservation laws with a
convex extension', {\it Proc. Nat. Acad. Sci.  U.S.A.} {\bf 68} (1971),
1686-1688.
\bibitem{Godunov61}
Godunov, S.K., 'An interesting class of quasilinear systems',
{\it Sov. Math. Dokl.},  {\bf 2} (1961), 947-949.
\end{thebibliography}
\end{document}